\documentclass[preprints,article,accept,moreauthors,pdftex]{mdpi}
\firstpage{1} 
\pubvolume{1}
\issuenum{1}
\articlenumber{0}
\pubyear{2022}
\copyrightyear{2022}
\datereceived{} 
\dateaccepted{} 
\datepublished{} 
\hreflink{https://doi.org/} 

\Title{Contribution to Excitonic Linewidth from Free Carrier--Exciton Scattering in Layered Materials: The example of h-BN}

\TitleCitation{Contribution to Excitonic Linewidth from Free Carrier--Exciton Scattering in Layered Materials: The example of h-BN}

\Author{M. F. C. Martins Quintela $^{1,2}$*\orcidA{} and N. M. R. Peres $^{1,2}$\orcidB{}}

\AuthorNames{M. F. C. Martins Quintela and N. M. R. Peres}

\AuthorCitation{Martins Quintela; M. F. C.; Peres, N. M. R.}

\address{%
$^{1}$ \quad Department of Physics and Physics Center of Minho and Porto Universities (CF--UM--UP), Campus of Gualtar, 4710-057, Braga, Portugal\\
$^{2}$ \quad International Iberian Nanotechnology Laboratory (INL), Av. Mestre Jos{\'e} Veiga, 4715-330, Braga, Portugal}

\corres{Correspondence: mfcmquintela@gmail.com}

\abstract{Scattering of excitons by free carriers is a phenomena which is especially important when considering moderately to heavily doped semiconductors in low temperature experiments, where the interaction of excitons with acoustic and optical phonons is reduced. In this paper, we consider the scattering of excitons by free carriers in monolayer hexagonal Boron Nitride encapsulated by a dielectric medium. We describe the excitonic states by variational wave functions, modeling the electrostatic interaction via the Rytova--Keldysh potential. Making the distinction between elastic and inelastic scattering, the relevance of each transition between excitonic states is also considered. Finally, we discuss the contribution of free carrier scattering to the excitonic linewidth, analyzing both its temperature and carrier density dependence.}


\begin{document}

\section{Introduction}

Scattering between excitons and free carriers has been observed experimentally since the 1960s in highly excited bulk semiconductors\cite{PhysRev.177.567,10.1063/1.1653814,10.1143/JPSJ.34.693,10.1002/pssb.2220580237,10.1002/pssa.2210220102,10.1002/pssb.2220710216,10.1002/pssb.2220780219,Levy1975ExperimentalIO}. In these bulk semiconductor systems, the scattering cross sections have been studied previously\cite{ELKOMOSS1977557,ELKOMOSS1979431}, with the distinction between elastic and inelastic scattering fundamental for the interpretation of the experimental data.

Recently, the quality of monolayer semiconductor samples has drastically increased\cite{Reynolds1986,Museur2008,Cadiz2017,Elias2019,Li2021a}, with advances in techniques such as molecular beam epitaxy\cite{Vuong2017}, chemical vapor deposition\cite{Kim2015} or solution methods\cite{Yano2000}. The improved quality of these samples allowed for a much more detailed study of excitons in mono- and few--layer materials, where well--defined resonance peaks are experimentally identifiable even at room temperature\cite{PhysRevLett.33.827,10.1063/1.93648,Poellmann2015,Merkl2019}. This then creates the necessity of calculating and measuring the excitonic linewidth. 

Several mechanisms contribute to the excitonic linewidth, such as acoustic\cite{PhysRevB.33.5512} and optical\cite{PhysRevB.34.2554} phonon scattering \cite{PhysRevB.103.235402}, radiative recombination\cite{10.1080/00018739100101522,10.1021/nl503799t}, as well as scattering in semiconductor alloys\cite{10.1063/1.94649,10.1063/1.102583}. Besides these mechanisms, others can play a part in determining the exciton linewidth. In this paper, we focus our attention on one of these mechanisms, namely scattering with free carriers. This scattering mechanism can play an important part in determining the excitonic linewidth, especially in systems with high density of free carriers and excitons\cite{10.1063/1.360578}, high pump fluences\cite{Djurii2005}, in tunneling experiments\cite{Young1990,Teran2009} or when an electric field is applied to the semiconductor\cite{10.1063/1.363972,10.1002/adom.202102132}.  Additionally, electron exciton scattering  processes also play an important role when studying exciton - polaron systems\cite{PhysRevB.103.075417}.

This paper is structured as follows. In Sec. \ref{sec:sec2}, we quickly review the approach outlined by Feng and Spector in Ref. \cite{FENG1987593} to the scattering between free carriers and excitons in semiconducting quantum wells. This derivation was performed in the central field \cite{Massey1951} and Born approximations \cite{Bates1953}.
In Sec. \ref{sec:sec3}, we turn our discussion to the total scattering cross section. We discuss the distinction between elastic and inelastic scattering, briefly reviewing the variational exciton wave functions for various states. We then explicitly compute the total cross section for a few select transitions, discussing the thresholds present in inelastic scattering processes. Finally, in Sec. \ref{sec:sec4}, we compute the contribution to excitonic linewidth from the scattering cross section with free carriers, analyzing its dependence on both the temperature of the system and the free carrier density in the monolayer.
\section{Free Carrier - Exciton Scattering\label{sec:sec2}}

In this section, we will follow the expressions derived by Feng and Spector in Ref. \cite{FENG1987593} for free carrier - exciton scattering, based on assuming two--dimensional (2D) gases of free carriers (electrons or holes) and excitons interacting with one another. The obtained expressions are derived following the same approach as those discussed by Mott and Massey for the general theory for tridimensional (3D) two--body collisions \cite{Mott1965}. The cross sections due to collisions between the carriers and excitons are then calculated using the central field \cite{Massey1951} and Born approximations \cite{Bates1953}.

\subsection{Differential Scattering Cross Section}

Let us begin by considering a two body system consisting of an exciton and a free carrier (electron or hole). The reduced mass of such a system is given by 
\begin{align}
	M &=\frac{m_c\left(m_e + m_h\right)}{m_c + m_e + m_h},
\end{align}
where $ m_{c/e/h} $ is the mass of the free carrier/ electron/hole. 

Following the derivations by Feng and Spector\cite{FENG1987593,FENG7096} in--depth in Appendix \ref{app:derivation}, the differential scattering cross section for our free carrier - exciton system is written as
\begin{equation}
	I_{fi}\left(\theta\right)=\frac{M^{2}}{2\pi\hbar^{4}k_{i}}\left|\int d^{2}\mathbf{r}\,d^{2}\mathbf{R}\,e^{i\mathbf{q}\cdot\mathbf{R}} V\left(\mathbf{r},\mathbf{R}\right)\chi_{f}^{\dagger}\left(\mathbf{r}\right)\chi_{i}\left(\mathbf{r}\right)\right|^{2},\label{eq:generic_crossection}
\end{equation}
where $ \mathbf{q}=\mathbf{k}_i-\mathbf{k}_f $ is the difference between the initial and final relative momentum of the system, $ \chi_{i/f} $ represent the initial/final exciton wave functions, and $ V\left(\mathbf{r},\mathbf{R}\right) $ the interaction potential between the free carrier and the exciton. In Eq. (\ref{eq:generic_crossection}), $ \mathbf{r} $ is the relative position vector of the electron and the hole in the exciton, and $ \mathbf{R} $ is the relative position vector from the free carrier to the center of mass of the exciton. 

The interaction potential between the free carrier and the exciton in the central field approximation\cite{Massey1951} will be modeled by the Rytova--Keldysh potential\cite{rytova1967,keldysh1979coulomb}, usually employed to describe excitonic phenomena in mono- and few--layer materials and obtained by solving the Poisson equation for a charge embedded in a thin film of vanishing thickness. In real space, the Rytova--Keldysh potential is given by
\begin{equation}
	V_{RK}(\mathbf{r})=\frac{\hbar c \alpha}{\epsilon}\frac{\pi}{2 r_{0}}\left[H_{0}\left(\epsilon\frac{r}{r_{0}}\right)-Y_{0}\left(\epsilon\frac{r}{r_{0}}\right)\right], \label{eq:RK}
\end{equation}
where $\alpha=1/137$ is the fine--structure constant, $\epsilon$ the mean dielectric constant of the medium above/below the layered material, $H_{0}\left(x\right)$ is the zeroth-order Struve function and $Y_{0}\left(x\right)$ is the zeroth-order Bessel function of the second kind. The parameter $r_{0}$ corresponds to an in--plane screening length related to the 2D polarizability of the material and can be calculated from the single particle Hamiltonian of the system \cite{PhysRevB.99.035429}. In the limit of zero screening length, the Rytova--Keldysh potential becomes the Coulomb potential. Considering, again, the interaction between the free carrier and the exciton, the interaction potential will be 
\begin{equation}\label{eq:pot_old}
	V\left(\mathbf{r},\mathbf{R}\right)=\pm \left[V_{RK}(\mathbf{r_{ch}})-V_{RK}(\mathbf{r_{ce}})\right],
\end{equation}
where $ \pm $ distinguishes between the free carrier being a hole ($ + $) or an electron ($ - $), $ r_{ch} $ is the distance between the free carrier and the hole of the exciton, and $ r_{ce} $ is the distance between the free carrier and the electron of the exciton. These two vectors can be written from $ \mathbf{r} $ and $ \mathbf{R} $ as \cite{FENG1987593}
\begin{align}\label{eq:vectors_from_masses}
	\mathbf{r}_{ch} &= \mathbf{R}-\frac{\sigma}{1+\sigma}\mathbf{r}, 
	&\mathbf{r}_{ce} &= \mathbf{R}+\frac{1}{1+\sigma}\mathbf{r},
\end{align}
where $ \sigma=m_e / m_h $ is the ratio between effective electron and hole masses. 
Returning to the discussion on the scattering cross section from Eq. (\ref{eq:generic_crossection}), the integration over $ R $ reads
\begin{equation}\label{eq:R_integration}
	\int d^{2}\mathbf{R}\,e^{i\mathbf{q}\cdot \mathbf{R}} V\left(\mathbf{r},\mathbf{R}\right)=\pm \left[\int d^{2}\mathbf{R}\,e^{i\mathbf{q}\cdot \mathbf{R}}V_{RK}(\mathbf{r_{ch}})-\int d^{2}\mathbf{R}\,e^{i\mathbf{q}\cdot \mathbf{R}}V_{RK}(\mathbf{r_{ce}})\right],
\end{equation}
which can be computed directly\cite{FENG7096} by performing a change of integration variables back to $\mathbf{r}_{ch},\,\mathbf{r}_{ce} $ and reads 
\begin{equation}\label{eq:change_variables_int}
	\pm\left[e^{iq\frac{\sigma}{1+\sigma}r\cos\left(\phi_{r}\right)}-e^{-iq\frac{1}{1+\sigma}r\cos\left(\phi_{r}\right)}\right]2\pi\frac{\hbar c\alpha}{\epsilon}\frac{1}{q\left(1+r_{0}q\right)},
\end{equation}
with $ \phi_{r} $ the angle between $ \mathbf{r} $ and $ \mathbf{q}$, and
\begin{equation}
	V_{RK}\left(\mathbf{q}\right)=2\pi\frac{\hbar c\alpha}{\epsilon}\frac{1}{q\left(1+r_{0}q\right)}
\end{equation}
the Fourier transform of the Rytova--Keldysh potential. Finally, the differential scattering cross section can be written as 

\begin{align}
	I_{fi}\left(\theta\right)=\frac{M^{2}}{2\pi\hbar^{4}k_{i}}\left|2\pi \frac{\hbar c\alpha}{\epsilon}\frac{1}{q\left(1+r_{0}q\right)}J\left(i\rightarrow f\right)\right|^{2}.\label{eq:cross-section_integrated_RK}
\end{align}
with the dependence on the initial and final exciton states included in $ J\left(i\rightarrow f\right) $, defined as
\begin{equation}\label{eq:J}
	J\left(i\rightarrow f\right)=\int_{0}^{+\infty}r\,dr\,\int_{0}^{2\pi} d\phi_{r}\,\left[e^{i\frac{\sigma}{1+\sigma}qr\cos \phi_{r}}-e^{-i\frac{1}{1+\sigma}qr\cos \phi_{r}}\right]\chi_{f}^{\dagger}\left(\mathbf{r}\right)\chi_{i}\left(\mathbf{r}\right).
\end{equation}

\section{Total Cross Section\label{sec:sec3}}

Knowing the differential cross section given by Eq. (\ref{eq:cross-section_integrated_RK}), we can now compute the full scattering cross section. To this effect, we must simply perform an angular integration in $ \theta $ as
\begin{align}\label{eq:full_scatter}
	Q_{\mathrm{i\rightarrow f}}&=\int_{-\pi}^{\pi}d\theta\,I_{\mathrm{i\rightarrow f}}\left(\theta\right).
\end{align}
Explicitly substituting Eq. (\ref{eq:cross-section_integrated_RK}), the full cross section is given by 
\begin{align}\label{eq:scatter_theta_RK_tot}
	Q_{\mathrm{i\rightarrow f}}\left(k_{i}\right)&=\frac{2\pi M^{2}\left(\hbar c\alpha\right)^{2}}{\hbar^{4}\epsilon^{2}k_{i}}\int_{-\pi}^{\pi}d\theta\,\left|\frac{J\left(i\rightarrow f\right)}{q(1+r_{0}q)}\right|^{2}.
\end{align}
To compute this integral, however, we must first define the exciton wave functions which we will consider when computing Eq. (\ref{eq:J}). We must also define the type of scattering in question, as it will introduce both the specific $ \theta $ dependence in $ \mathbf{q} $ as well as specific thresholds for the relative momentum of the free carrier - exciton system from conservation of energy.

\subsection{Elastic Scattering}

In an elastic scattering process, the exciton remains in its ground state after the collision, meaning $ \left|\mathbf{k}_i\right|=\left|\mathbf{k}_f\right| $. As such, we can write
\begin{equation}
	q= 2 \left|\sin\left(\frac{\theta}{2}\right)\right| k_i.\label{eq:elast_q}
\end{equation}
Additionally, we also have $ \chi_{f,i}\left(\mathbf{r}\right)=\chi_{1s}\left(r\right) $, where we consider a simple variational \emph{ansatz} \cite{Quintela,Gomes2021} based on the eigenfunctions of the two--dimensional Hydrogen atom\cite{Lee1979,PhysRevA.43.1186} and given by 
\begin{equation}\label{eq:ansatz}
	\chi_{1s}\left(r\right)=\mathcal{N}_{1s}e^{-r\gamma_{1s}/2},
\end{equation}
with $ \mathcal{N}_{1s} $ a normalization constant given by 
\begin{equation}\label{eq:normalization}
	\mathcal{N}_{1s}=\left[\int rdrd\theta\,\left(e^{-r\gamma_{1s}/2}\right)^{2}\right]^{-1/2}=\frac{\gamma_{1s}}{\sqrt{2\pi}}
\end{equation}
and $ \gamma_{1s} $ a variational parameter. This variational parameter is computed by minimization of the energy expectation value of the Wannier equation\cite{Wannier1937}
\begin{equation}\label{eq:wannier}
	H=-\frac{\hbar^{2} c^{2}}{2\mu}\nabla^2+V_{RK}(\mathbf{r}),
\end{equation}
with $ 	V_{RK}(\mathbf{r}) $ the Rytova--Keldysh potential.

With this \emph{ansatz}, we can directly substitute the wave function into $ J\left(i\rightarrow f\right) $, given by Eq. (\ref{eq:J}), and obtain
\begin{align}
	J_{\mathrm{elast}}\left(q\right)=J\left(\mathrm{1s}\rightarrow \mathrm{1s}\right)
	&=\gamma_{1s}^{3}\left[\frac{1}{\left[\left(\frac{\sigma}{1+\sigma}q\right)^{2}+\gamma_{1s}^{2}\right]^{3/2}}-\frac{1}{\left[\left(\frac{1}{1+\sigma}q\right)^{2}+\gamma_{1s}^{2}\right]^{3/2}}\right]\label{eq:J_transition_elastic_exponential}
\end{align}
after integration. This is then substituted into Eq. (\ref{eq:scatter_theta_RK_tot}), reading 
\begin{align}\label{eq:scatter_theta_RK_tot_elast}
	Q_{\mathrm{elast}}\left(k_{i}\right)&=\frac{2\pi M^{2}\left(\hbar c\alpha\right)^{2}}{\hbar^{4}\epsilon^{2}k_{i}}\int_{-\pi}^{\pi}d\theta\,\left|\frac{J_{\mathrm{elast}}\left(q\right)}{q(1+r_{0}q)}\right|^{2},
\end{align}
where $ q $ is given by Eq. (\ref{eq:elast_q}). This integral has no analytical solution and must be computed numerically.

\begin{figure}
	\centering{}\includegraphics{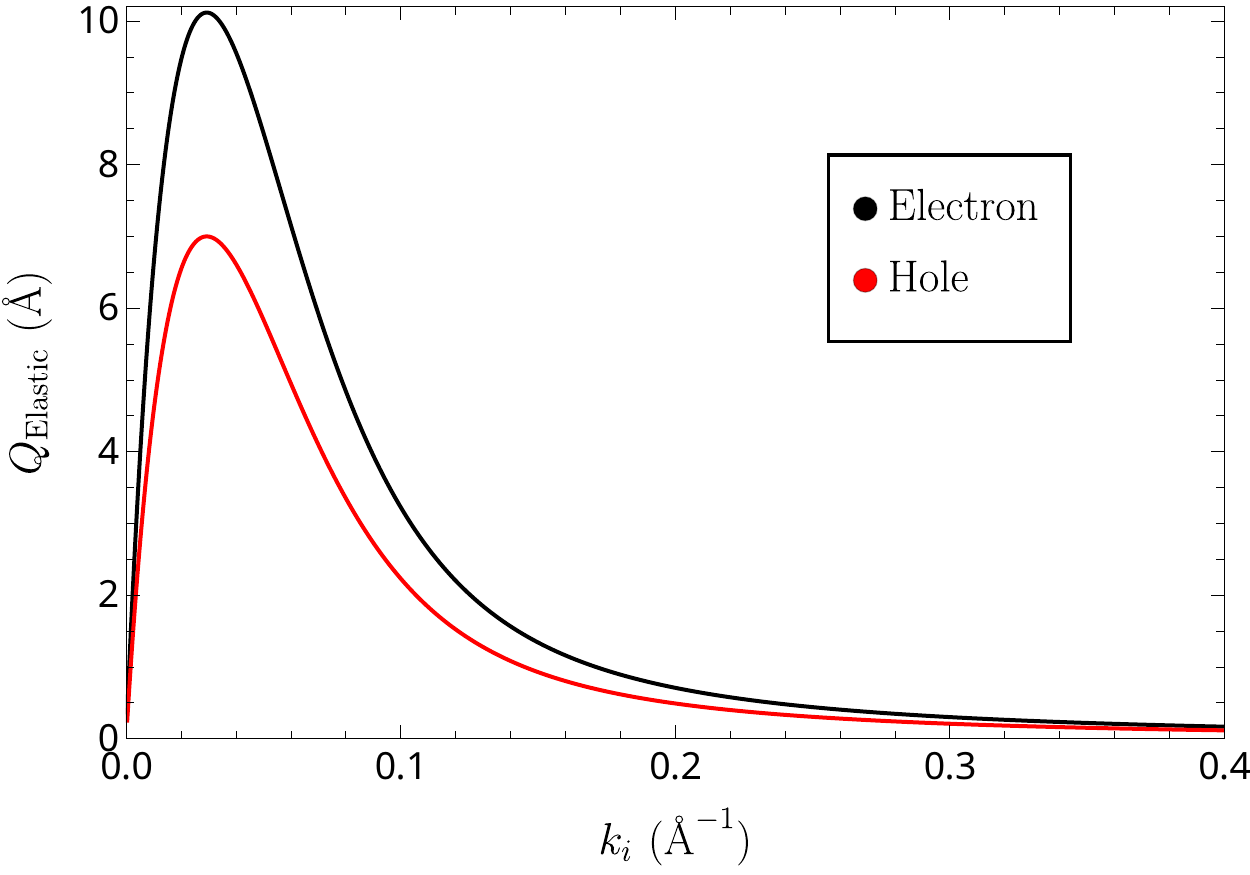}
	\caption{Total elastic cross section for electron--exciton (black) and hole--exciton (red) scattering in $ \mathrm{hBN} $ encapsulated in fused quartz as a function of the initial relative wave vector. \label{fig:cross-section_coulomb_GaAs}}
\end{figure}

To finalize the computation of the elastic cross section, we must choose a set of material specific parameters. We consider those corresponding to monolayer hexagonal Boron--Nitride ($ \mathrm{hBN} $) encapsulated in fused quartz, with dielectric constant $ \epsilon=3.8 $ \cite{Serway2003}. The electron and hole masses in this material are $ m_e = 0.83\, m_0$, $ m_h = 0.63\, m_0$ \cite{Ferreira19}, with $ m_0 $ the electron rest mass, and $ r_0 = 10\,\text{\AA}$ \cite{henriques_optical_2020}. The obtained variational energy for the $ 1s $ excitonic state is $ E_{1s}=-58.9\,\mathrm{meV} $.

Varying the initial relative wave vector, we obtain the plot of the total elastic cross section from Eq. (\ref{eq:scatter_theta_RK_tot}) in Fig. (\ref{fig:cross-section_coulomb_GaAs}). We can see that the cross section for electron scattering is always larger than that for hole scattering, as expected from the fact that the reduced mass of the system is larger when the free carrier considered is an electron. A very quick increase from zero relative momentum up to a global maximum is also observed, consistent with the results of Feng and Spector \cite{FENG7096} for elastic scattering.

\subsection{Inelastic Scattering}

Let us now consider inelastic scattering between free carriers and the exciton. The relative momenta is now given by
\begin{equation}\label{eq:momenta_1}
	q^{2} = k_{f}^{2} + k_{i}^{2}-2k_{i}k_{f}\cos(\theta), 
\end{equation}
with $ k_{f} $ obtained from conservation of energy as 
\begin{equation}\label{eq:conservation_energy}
	k_{f}^{2}=k_{i}^{2}-\frac{2M}{\hbar^{2}}(E_{f}-E_{i}).
\end{equation}
Here, $ E_{f/i} $ is the energy of the final/initial state of the exciton, respectively. Substituting this relation into Eq. (\ref{eq:momenta_1}), we obtain
\begin{equation}\label{eq:momenta_2}
	q^{2} = 2k_{i}^{2}-\frac{2M}{\hbar^{2}}\Delta_{f,i}-2k_{i}^{2}\sqrt{1-\frac{2M}{\hbar^{2}}\frac{\Delta_{f,i}}{k_{i}^{2}}}\cos(\theta),
\end{equation}
with $ \Delta_{f,i}=E_{f}-E_{i} $. A threshold in $ k_i $ below which no scattering is allowed is immediately evident, obtained from Eq. (\ref{eq:momenta_2}) as 
\begin{equation}\label{eq:threshold}
	k_{\mathrm{min}}=\sqrt{\frac{2M}{\hbar^{2}}(E_{f}-E_{i})}.
\end{equation}
Below this threshold, there is not enough energy in the scattering process to allow the jump between excitonic states $ i\rightarrow f $. 

\begin{figure}
	\centering{}\includegraphics{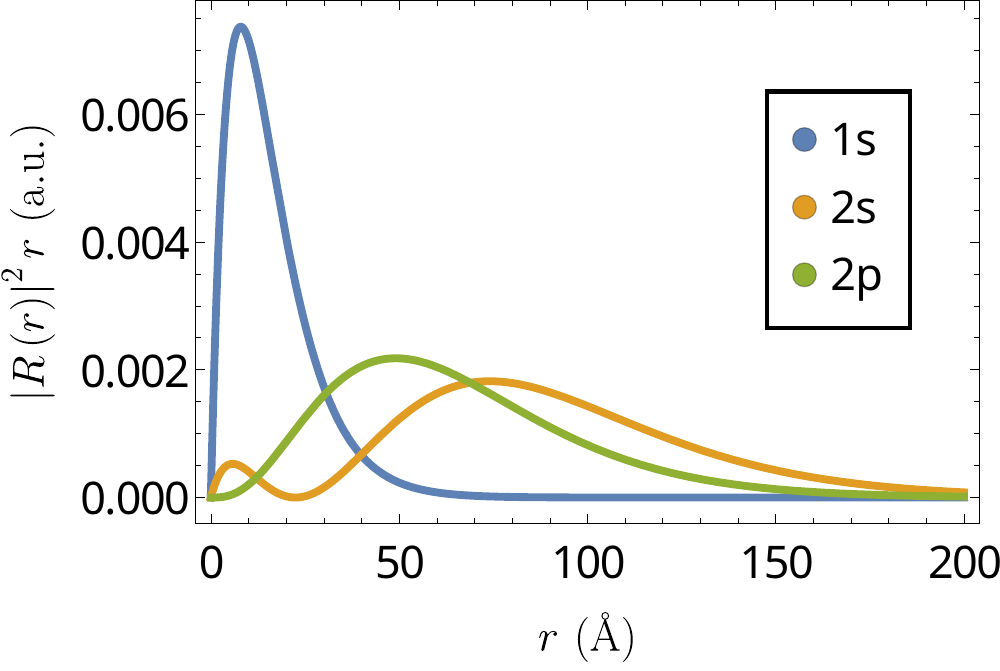}
	\caption{Radial probability density for the $ 1s $ (blue), $ 2s $ (orange) and $ 2p $ (green) states as described by the variational wave functions of Eq. (\ref{eq:ansatz}) and Eq. (\ref{eq:2s_2p_ansatz}) in a $ \mathrm{hBN} $ monolayer encapsulated in fused quartz. \label{fig:wave_functions}}
\end{figure}

Besides knowing the energies of the final states, we must also know their wave functions. These are obtained\cite{Lee1979,Quintela,Gomes2021} in a similar form to Eq. (\ref{eq:ansatz}) and are given by 
\begin{align}\label{eq:2s_2p_ansatz}
	\chi_{2s}\left(r\right)&=\mathcal{N}_{2s}\left(1-\frac{r}{d}\right)e^{-r\gamma_{2s}/2}, &
	\chi_{2p_{\pm}}\left(r\right)&=\mathcal{N}_{2p}r e^{\pm i \theta}e^{-r\gamma_{2p}/2},
\end{align}
where $ \gamma_{2s},\gamma_{2p} $ are variational parameters, $ \mathcal{N}_{2s},\mathcal{N}_{2p} $ are normalization constant given by 
\begin{align}
	\mathcal{N}_{2s}&=\frac{2\gamma_{2s}^{2}}{\sqrt{\pi}\sqrt{3\gamma_{1s}^{2}-2\gamma_{1s}\gamma_{2s}+3\gamma_{2s}^{2}}},&
	\mathcal{N}_{2p}&=\frac{\gamma_{2p}^{2}}{2\sqrt{3\pi}},
\end{align}
and $ d $ is a parameter obtained by imposing orthogonality between $ \chi_{1s} $ and $ \chi_{2s} $, given by 
\begin{equation}
	d=\frac{4}{\gamma_{1s}+\gamma_{2s}}.
\end{equation}
These wave functions, together with the $ 1s $ wave function, are plotted in Fig. (\ref{fig:wave_functions}) for monolayer $ \mathrm{hBN} $ encapsulated in fused quartz.

\subsubsection{$ 1s \rightarrow 2s $ Transitions}
We will first consider $ 1s \rightarrow 2s $ transitions. To compute $ J_{2s}=J\left(1s\rightarrow 2s\right) $, we recall Eq. (\ref{eq:J}) and, after integration, obtain
\begin{align}\label{eq:J_1s2s}
	J_{2s}
	&= \frac{3\gamma_{1s}\left(\gamma_{1s}+\gamma_{2s}\right)\gamma_{2s}^{2}}{\sqrt{6\gamma_{1s}^{2}-4\gamma_{1s}\gamma_{2s}+6\gamma_{2s}^{2}}} \left[\frac{q^{2}\left(\frac{\sigma}{1+\sigma}\right)^2}{\left[q^2\left(\frac{\sigma}{1+\sigma}\right)^2+\left(\frac{\gamma_{1s}+\gamma_{2s}}{2}\right)^{2}\right]^{5/2}}-\frac{q^{2}\left(\frac{1}{1+\sigma}\right)^2}{\left[q^2\left(\frac{1}{1+\sigma}\right)^2+\left(\frac{\gamma_{1s}+\gamma_{2s}}{2}\right)^{2}\right]^{5/2}}\right],\nonumber
\end{align}
where $ q $ is obtained from Eq. (\ref{eq:momenta_2}) as 
\begin{equation}\label{eq:q_1s2s}
	q^{2}=2k_{i}^{2}-\frac{2M}{\hbar^{2}}\Delta_{2s,1s}-2k_{i}^{2}\sqrt{1-\frac{2M}{\hbar^{2}}\frac{\Delta_{2s,1s}}{k_{i}^{2}}}\cos(\theta).
\end{equation}

As discussed above, the energy of the $ 2s $ state is obtained by minimization of the Wannier equation with the variational wave functions and its value is $ E_{2s}=-8.83\,\mathrm{meV} $ for our system. Explicitly computing the thresholds from Eq. (\ref{eq:threshold}), we obtain $ k_{\mathrm{min}}= 0.0833\,\text{\AA}^{-1}$ for electron--exciton scattering and $ k_{\mathrm{min}}= 0.0760\,\text{\AA}^{-1}$ for hole--exciton scattering.

\subsubsection{$ 1s \rightarrow 2p $ Transitions}

For computing $ J\left(1s\rightarrow 2p_{\pm}\right) $ following Eq. (\ref{eq:J}), we must quickly consider the distinction between $ p_{\pm} $ states. This is, however, not important, as the two states are degenerate and the two integrals $ J\left(1s\rightarrow 2p_{+}\right) $, $ J\left(1s\rightarrow 2p_{-}\right) $ are, in fact, equal. As such, we take into account the two $ 2p $ states by multiplying the total cross section by an angular momentum degeneracy factor $ g_{\ell}=2 $.

Explicitly, $ J_{2p}=J\left(1s\rightarrow 2p\right) $ is given by 
\begin{align}\label{eq:J_1s2p}
	J_{2p}
	&=\frac{3\gamma_{1s}\gamma_{2p}^{2}\left(\gamma_{1s}+\gamma_{2p}\right)}{2\sqrt{6}}\left[\frac{iq\frac{\sigma}{1+\sigma}}{\left[q^2\left(\frac{\sigma}{1+\sigma}\right)^2+\left(\frac{\gamma_{1s}+\gamma_{2p}}{2}\right)^{2}\right]^{5/2}}+\frac{iq\frac{1}{1+\sigma}}{\left[q^2\left(\frac{1}{1+\sigma}\right)^2+\left(\frac{\gamma_{1s}+\gamma_{2p}}{2}\right)^{2}\right]^{5/2}}\right].\nonumber
\end{align}
The relative momentum $ q $ is defined analogously to Eq. (\ref{eq:q_1s2s}), with the only factor missing being the energy of the $ 2p_{\pm} $ states.

For our system, this energy is $ E_{2p}=-10.2\,\mathrm{meV} $. As such, the thresholds from Eq. (\ref{eq:threshold}) are now $ k_{\mathrm{min}}= 0.0822\,\text{\AA}^{-1}$ for electron--exciton scattering and $ k_{\mathrm{min}}=0.0749\,\text{\AA}^{-1}$ for hole--exciton scattering. 

\subsection{Joint Elastic and Inelastic Scattering}

\begin{figure}
	\centering{}\includegraphics{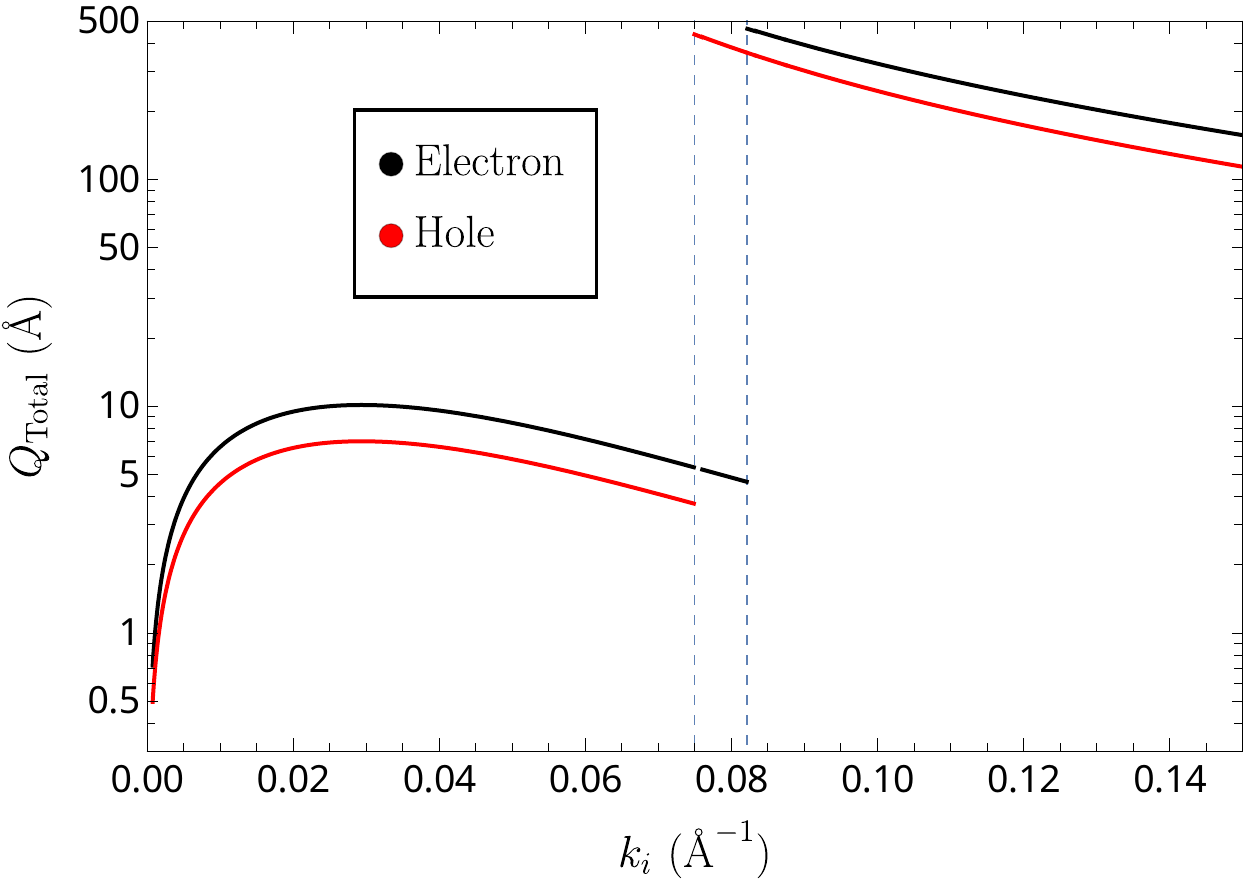}
	\caption{Total cross section for electron--exciton (black) and hole--exciton (red) scattering in $ \mathrm{hBN} $ encapsulated in fused quartz as a function of the initial relative wave vector. Vertical dashed lines represent the electron--exciton and hole--exciton scattering thresholds for the $ 1s \rightarrow 2p $ transition, located at $ k_{\mathrm{min},e^{-}}= 0.0822\,\text{\AA}^{-1}$ and $ k_{\mathrm{min},h}= 0.0749\,\text{\AA}^{-1}$.
		\label{fig:cross-section_total}}
\end{figure}

Finally, we consider the joint contribution to the scattering cross section from both elastic and inelastic scattering. This cross section will, therefore, involve a sum over final states, where only the $ 1s\rightarrow 1s $ contribution originates from elastic scattering processes. Explicitly, and for an arbitrary set of final exciton states $ f $, the total scattering cross section is given by
\begin{equation}\label{eq:scatt_total_1}
	Q_{\mathrm{Total}}=\sum_{f}Q_{1s\rightarrow f}.
\end{equation}
For the three transitions discussed previously, the sum in Eq. (\ref{eq:scatt_total_1}) is restricted and is explicitly written as 
\begin{equation}\label{eq:scatt_total}
	Q_{\mathrm{Total}}=Q_{1s\rightarrow 1s}+Q_{1s\rightarrow 2s}+Q_{1s\rightarrow 2p}.
\end{equation}
This total scattering cross section is plotted in Fig. (\ref{fig:cross-section_total}) for both types of free carriers, together with the dashed lines representing the thresholds for the inelastic scattering processes considered.

Analyzing Fig. (\ref{fig:cross-section_total}), we can see that, when the same scattering process is allowed for both types of free carriers, electron--exciton scattering has a cross section roughly $ 1.5\times $ larger. This trend is, however, inverted between the threshold momentum for hole--exciton $ 1s \rightarrow 2p $ scattering and the threshold momentum for electron--exciton $ 1s \rightarrow 2p $, \emph{i.e.}, between $ k_i = 0.0749\,\text{\AA}^{-1} $ and $ k_i = 0.0822\,\text{\AA}^{-1}$. In this momentum range, the dominant $ 1s \rightarrow 2p $ process is already allowed for hole--exciton scattering, leading to a vastly superior cross section relative to that for electron--exciton scattering.

The final threshold included, visible in Fig. (\ref{fig:cross-section_total}) as the dashed purple lines, originates from $ 1s \rightarrow 3d $ scattering, as $ E_{3d}=-3.74\,\mathrm{meV} $ is the lowest energy state after $ 2s $. These take place at $ k_{\mathrm{min}}=0.0875\,\text{\AA}^{-1} $ for electron--exciton scattering and $ k_{\mathrm{min}}=0.0798\,\text{\AA}^{-1} $ for hole--exciton scattering. These are, however, much smaller than the peaks in Fig. (\ref{fig:cross-section_total}), similarly to what is observed Ref. \cite{FENG1987593}, and are invisible in Fig. (\ref{fig:cross-section_total}). 

\section{Scattering Contribution to Exciton Linewidth\label{sec:sec4}}

To conclude our study of the scattering of excitons with free carriers in layered materials, we will now discuss the contribution from these scattering processes to the excitonic linewidth. This will provide a point of comparison against experimental studies\cite{Selig2016}. Although other phenomena will also contribute to this linewidth, such as radiative lifetimes\cite{Palummo2015} and phonon scattering\cite{Henriques2021}, the dependence of the free carrier scattering on both temperature and carrier density should provide a good distinction of the various contributing processes. 

The contribution to the excitonic linewidth from free carrier scattering is given by\cite{PhysRevB.34.2554,FENG1987459,feng1989excitons,10.1063/1.364274}
\begin{equation}\label{eq:linewidth}
	\Gamma_{\mathrm{Total}}=\sum_{f}\frac{2\hbar^{2}}{\pi M}\int_{0}^{\infty}dk\,k^{2}f\left(\frac{m_c+m_e+m_h}{m_e+m_h}k\right)Q_{1s\rightarrow f},
\end{equation}
where $ Q_{1s\rightarrow f} $ is, as described earlier, the scattering cross section associated with a specific transition from the excitonic ground state to a final state $ f $. As we are summing over final states $ f $ and only $ Q_{1s\rightarrow f} $ depends on the final state, this is equivalent to switching the sum over final states and the integral and writing 
\begin{equation}\label{eq:linewidth_tot}
	\Gamma_{\mathrm{Total}}=\frac{2\hbar^{2}}{\pi M}\int_{0}^{\infty}dk\,k^{2}n_{F}\left(\frac{m_c+m_e+m_h}{m_e+m_h}k\right)Q_{\mathrm{Total}},
\end{equation}
where $ Q_{\mathrm{Total}} $ is the total scattering cross section as plotted in Fig. (\ref{fig:cross-section_total}).
Here, $ n_{F}(k) $ is the Fermi--Dirac distribution for free carriers, given by 
\begin{equation}\label{eq:fermi_dirac-distr}
	n_{F}(k)=\frac{1}{e^{\frac{E_k-E_F}{k_B T}}+1}
\end{equation}
where the dispersion relation is given by
\begin{equation}\label{eq:free-carrier_dispersion}
	E_k=\frac{\hbar^{2}}{2 m_{c}}k^{2}
\end{equation}
and the Fermi energy is
\begin{equation}\label{eq:fermi_energy_free}
	E_F=2\pi \frac{\hbar^{2}}{2 m_{c}}n,
\end{equation}
with $ n $ the area density of free carriers. We consider carrier densities up to a maximum of $ 10^{12}\,\mathrm{cm}^{-2} $, where the average separation of free carriers $ d_{c}= 2 (\pi n)^{-1/2} $ is still larger but already of the order of the root mean square (RMS) exciton radius, given by \begin{equation}\label{eq_rms}
	r_{\mathrm{RMS};n}=\int_{0}^{\infty}\int_{0}^{2\pi} rdrd\theta\,\psi_{n}(r,\theta)^{\dagger} r^{2} \psi_{n}(r,\theta).
\end{equation}
For the two excitonic states most relevant to the scattering cross section, $ 1s $ and $ 2p $, the RMS exciton radius is $ r_{\mathrm{RMS};1s}=19.5\,\text{\AA} $ and $ r_{\mathrm{RMS};2p}=73.2\,\text{\AA} $, respectively, while the average separation between free carriers is $ d_{c}=113\,\text{\AA} $.

\begin{figure}
	\includegraphics[scale=0.78]{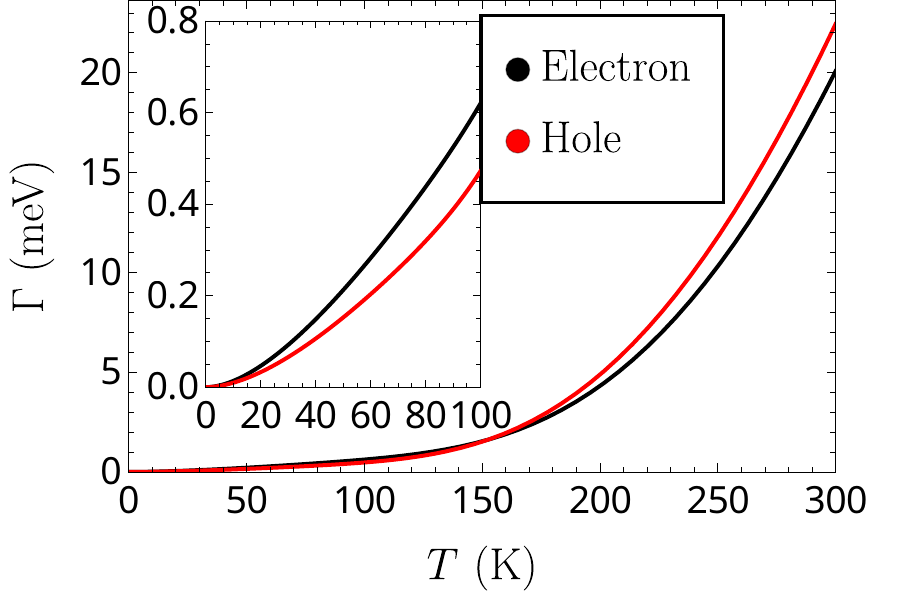}\includegraphics[scale=0.78]{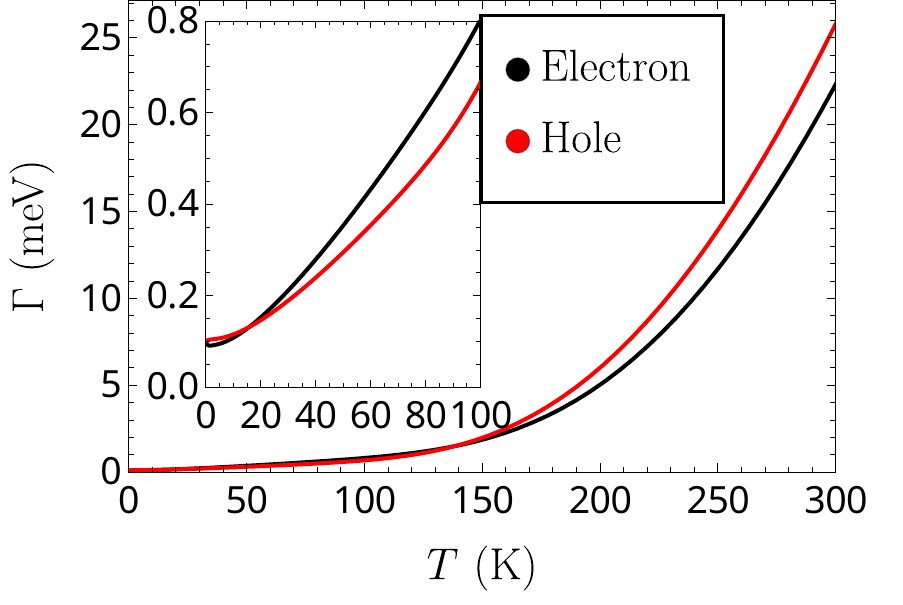}
	\caption{Contribution to exciton linewidth from scattering with free carriers as a function of the temperature at constant free carrier area density $n=10^{9}\,\text{cm}^{-2}$ (left) and $n=10^{12}\,\text{cm}^{-2}$ (right). Insets show the low--temperature region of the plot for clarity.\label{fig:linewidth_const_density}}
\end{figure}

\begin{figure}
	\centering{}\includegraphics[scale=0.9]{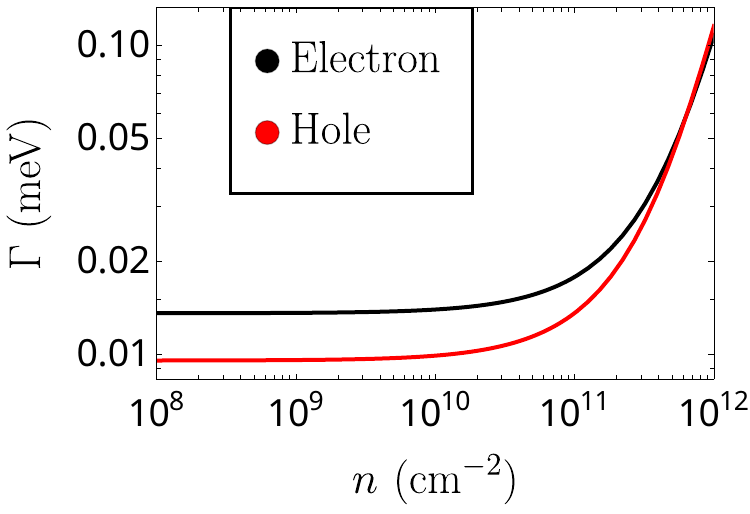}\includegraphics[scale=0.9]{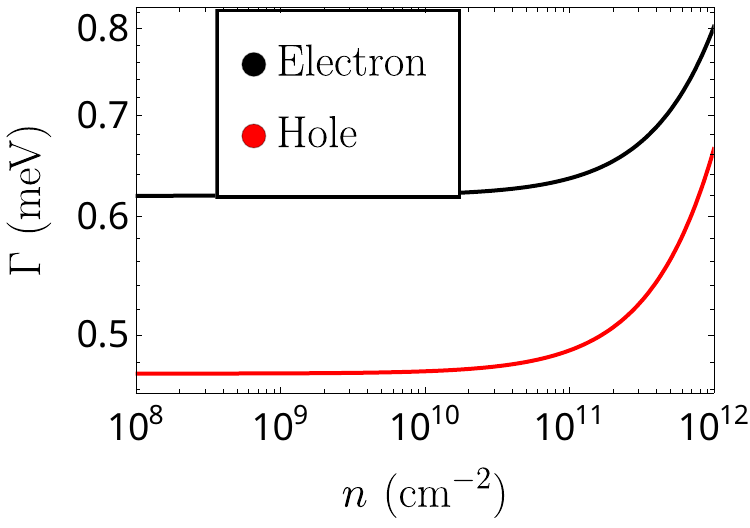}
	\includegraphics[scale=0.9]{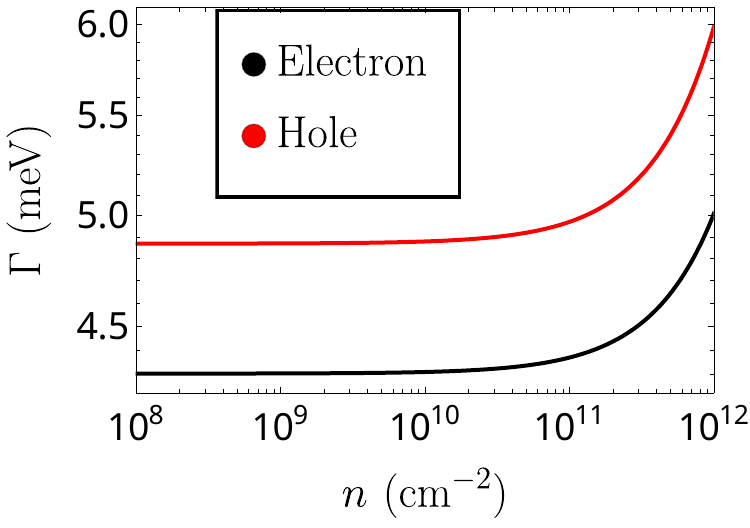}\includegraphics[scale=0.9]{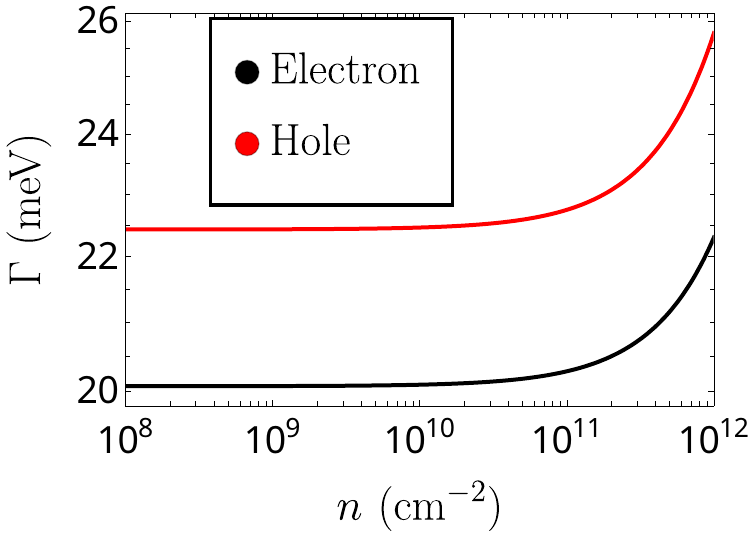}
	\caption{Contribution to exciton linewidth from scattering with free carriers as a function of the free carrier area density at constant temperature $ T=10\,\mathrm{K}$ (top-left), $  T=100\,\mathrm{K}$ (top-right), $  T=200\,\mathrm{K}$ (bottom-left), and $  T=300\,\mathrm{K}$ (bottom-right). \label{fig:linewidth_const_temp}}
\end{figure}

As before, we must compute the integral of Eq. (\ref{eq:linewidth_tot}) numerically. The specific methodology for the discretization of Eq. (\ref{eq:linewidth_tot}) is discussed in Sec. \ref{app:numeric_integral}. Choosing a Gauss--Legendre quadrature\cite{Kythe2002} of size $ N=450 $, the results for the contribution of scattering with free carriers to the excitonic linewidth are presented in Fig. (\ref{fig:linewidth_const_density}) as a function of temperature for free carrier area densities of $ n = 10^{9}\,\,\text{cm}^{-2} $ and $ n = 10^{12}\,\,\text{cm}^{-2} $. In Fig. (\ref{fig:linewidth_const_temp}), we present the excitonic linewidth as a function of the free carrier area density for four distinct values of the temperature $ T $ between $ 10\,\mathrm{K} $ and $ 300\,\mathrm{K}  $. 
\section{Conclusions}

In this paper, we studied the effects of scattering between free carriers and excitons in monolayer materials and its contribution to the excitonic linewidth. To this end, we began by reviewing the general form of the expressions for the differential cross section between free carriers and excitons in two dimensions\cite{FENG1987593}, as well as the inclusion of screening in the differential cross section. Quickly reviewing the computation of variational functions for the exciton wave functions, we discussed both elastic and inelastic scattering processes. 

With the differential cross section known for both elastic and inelastic scattering, we proceeded to the computation of the total scattering cross section. 
While elastic scattering is the sole contributor for low relative momentum, inelastic $ 1s\rightarrow 2p $ scattering dominates the total cross section after it becomes allowed. This dominant scattering has a maximum cross section roughly $ 40 $ times larger than that of elastic scattering, and at least $ 400 $ times larger than both $ 1s\rightarrow 2s $ and $ 1s\rightarrow 3d $ inelastic scatterings. This dominant behavior of $ 1s\rightarrow 2p $ transitions is very similar to what is is presented in Figs. (2), (9) of Ref. \cite{FENG1987593}. This behavior is, of course, dependent on the ratio between electron and hole masses, although we only focus our attention on the masses characteristic of single layer hBN. 

After discussing the relation between the total scattering cross section and the excitonic linewidth, we considered both its dependence on both temperature and free carrier density. First looking at the exciton linewidth for fixed values of the carrier density, we observe that the difference of linewidth between the two densities studied ($ 10^{9}\,\,\text{cm}^{-2} $ and $ 10^{12}\,\,\text{cm}^{-2} $) is noticeable in both the finite value of the linewidth as $ T$ approaches $0\,\mathrm{K} $ and in the crossing of the two curves happening slightly earlier at higher carrier density (at $ T\gtrsim150\,\mathrm{K} $ for $ n=10^{9}\,\,\text{cm}^{-2} $ and at $ T\approx140\,\mathrm{K} $ for $ n=10^{12}\,\,\text{cm}^{-2} $). An even greater noticeable change in behavior would be present at densities above $ 10^{13}\,\,\text{cm}^{-2} $, reasonably larger than those usually obtained in experimental works. These higher carrier densities would already imply an average free carrier separation $ d_{c} $ much smaller than the RMS exciton radius for the $ 2p $ excitonic state, which would make more complex excitonic phenomena, such as biexcitons and trions\cite{Lampert1958,Kheng1993,Chen2002,Li2003}, increasingly significant.

Regarding the excitonic linewidth at fixed temperatures, the free carrier scattering contribution remained in the $ 0.5-20\,\mathrm{meV} $ range for temperatures between $ 100-300\,\mathrm{K} $, although a much faster increase for higher densities is observed at $ T=10\,\mathrm{K} $. For temperatures in the range $ 100-300\,\mathrm{K} $, the computed linewidth remained essentially constant as carrier density increases until roughly $ n\approx10^{11}\,\,\text{cm}^{-2} $, although the actual value of the linewidth is strongly dependent on the temperature. Past $ n\approx10^{11}\,\,\text{cm}^{-2} $, the computed linewidth rapidly increases for all values of the temperature, although the growth occurs sooner for lower temperatures. It should, therefore, be feasible to measure this contribution to the excitonic linewidth, as this is both an attainable carrier density \cite{Du2016,Kuiri2021,Li2021}, and good quality samples of both hBN and transition metal dichalcogenides have been grown in the past which presented excitonic linewidths in this order of magnitude\cite{Reynolds1986,Museur2008,Cadiz2017,Elias2019,Li2021a}.


\funding{M. F. C. M. Q. acknowledges the International Nanotechnology Laboratory (INL) and the Portuguese Foundation for Science and Technology (FCT) for the Quantum Portugal Initiative (QPI) grant SFRH/BD/151114/2021. 
N. M. R. P. acknowledges support by the Portuguese Foundation for Science and Technology (FCT) in the framework of the Strategic Funding UIDB/04650/2020, COMPETE 2020, PORTUGAL 2020, FEDER, and  FCT through projects PTDC/FIS-MAC/2045/2021, EXPL/FIS-MAC/0953/ 2021, and from the European Commission through the project Graphene Driven Revolutions in ICT and Beyond (Ref. No. 881603, CORE 3).}

\dataavailability{Not applicable.} 

\conflictsofinterest{The authors declare no conflict of interest.} 


\abbreviations{Abbreviations}{
The following abbreviations are used in this manuscript:\\

\noindent 
\begin{tabular}{@{}ll}
hBN & hexagonal Boron Nitride\\
RMS & root mean squared
\end{tabular}
}

\appendixtitles{yes} 
\appendixstart
\appendix
\section{Derivation of Differential Scattering Cross Section \label{app:derivation}}

For the derivation of the differential scattering cross section of Eq. (\ref{eq:generic_crossection}), we will follow closely the derivation performed by Feng and Spector \cite{FENG7096}. We will consider an encounter between two bodies $ A $ (free carrier) and $ B $ (exciton) which are in their ground state prior to the collision. The reduced mass of this system is $ M=\frac{m_A m_B}{m_A + m_B} $. Prior to the collision, the internal motions of each of the two bodies are given by their independent Hamiltonians
\begin{equation}\label{key}
	\begin{aligned}
	H_{A}\chi_{A,n}\left(\mathbf{r}_{A}\right)&=E_{A,n}\chi_{A,n}\left(\mathbf{r}_{A}\right),&
	H_{B}\chi_{B,m}\left(\mathbf{r}_{B}\right)&=E_{B,m}\chi_{B,m}\left(\mathbf{r}_{B}\right),
	\end{aligned}
\end{equation}
where $ \chi_{A,n}/\chi_{B,m} $ are the independent wave functions for the internal motion of the two bodies $ A/B $ in the states $ n/m $, respectively, and $ E_{A,n}/E_{B,m} $ their energies. In the absence of interaction of the two bodies, their relative motion is given by 
\begin{equation}\label{eq:wave_absence}
	\left[-\frac{\hbar^{2}}{2M}\nabla^{2}-\frac{1}{2} M \mathbf{v}^{2}\right] F_{11}(\mathbf{R})=0,
\end{equation}
with $ \mathbf{v} $ and $ \mathbf{R} $ the relative velocity and position of the two particles and $  F_{11}(\mathbf{R}) $ the part of the wave function related to the relative motion of the particles in their ground state. 

The complete wave equation for this system is given by 
\begin{equation}
		\left[-\frac{\hbar^{2}}{2M}\nabla^{2}-\frac{1}{2} M \mathbf{v}^{2}+H_{A}-E_{A,1}+H_{B}-E_{B,1}+V\left(\mathbf{R}, \mathbf{r}_{A}, \mathbf{r}_{B}\right)\right] \Psi\left(\mathbf{R}, \mathbf{r}_{A}, \mathbf{r}_{B}\right)=0,
\end{equation}
where $ V\left(\mathbf{R}, \mathbf{r}_{A}, \mathbf{r}_{B}\right) $ is the interaction potential between the two particles. The wave function can then be expanded in terms of the basis functions $ \chi_{A,n}/\chi_{B,m} $, reading 
\begin{equation}
	\Psi\left(\mathbf{R}, \mathbf{r}_{A}, \mathbf{r}_{B}\right)=\sum_{n, m} F_{n m}(\mathbf{R}) \chi_{A ,n}\left(\mathbf{r}_{A}\right) \chi_{B ,m}\left(\mathbf{r}_{B}\right),
\end{equation}
meaning that $ F_{n m}(\mathbf{R}) $ must obey
\begin{equation}\label{eq:Fnm_eq}
	\left[\nabla^{2}+k^{2}\right] F_{n m}(\mathbf{R})  =\frac{2 M}{\hbar^{2}}\int d\mathbf{r}_{A} d\mathbf{r}_{B} \,V\left(\mathbf{r}_{A}, \mathbf{r}_{B}, \mathbf{R}\right) \Psi\left(\mathbf{r}_{A}, \mathbf{r}_{B}, \mathbf{R}\right) \chi_{A n}^{*}\left(\mathbf{r}_{A}\right) \chi_{B m}^{*}\left(\mathbf{r}_{B}\right),
\end{equation}
where 
\begin{equation}\label{eq:k2}
	\mathbf{k}^{2}=\frac{2 M}{\hbar^{2}}\left[\frac{1}{2} M \mathbf{v}^{2}+E_{A,1}-E_{A,n}+E_{B,1}-E_{B,m}\right].
\end{equation}
The relation in Eq. (\ref{eq:k2}) immediately leads to the thresholds for inelastic scattering, as $ k $ being real implies $ \frac{1}{2} M \mathbf{v}^{2}>E_{A,n}-E_{A,1}+E_{B,m}-E_{B,1} $.

The solution to Eq. (\ref{eq:Fnm_eq}) reads
\begin{equation}
	F_{n m}(\mathbf{R})=\frac{-i M}{2\hbar^{2}} \int d\mathbf{r}_{A} d\mathbf{r}_{B}d\mathbf{R}^{\prime}\,V\left(\mathbf{r}_{A}, \mathbf{r}_{B}, \mathbf{R}^{\prime}\right)\Psi\left(\mathbf{R}^{\prime}, \mathbf{r}_{A}, \mathbf{r}_{B}\right) \chi_{A ,n}^{*}\left(\mathbf{r}_{A}\right) \chi_{B, m}^{*}\left(\mathbf{r}_{B}\right) \mathrm{H}_{0}^{1}\left(k\left|\mathbf{R}-\mathbf{R}^{\prime}\right|\right),
\end{equation}
where $ \mathrm{H}_{0}^{1}\left(k\left|\mathbf{R}-\mathbf{R}^{\prime}\right|\right) $ is the Hankel function of the first kind which is the solution to Eq. (\ref{eq:wave_absence}). This function satisfies the boundary condition that, for $ \mathbf{R}\gg\mathbf{R}^{\prime} $, the solution represents an outgoing circular wave. 

Following in identical procedure to that which is used to apply the Born approximation in a 3D system, the asymptotic regime as $ \mathbf{R}\rightarrow\infty $ for $ F_{n m}(\mathbf{R}) $ reads 
\begin{equation}\label{eq:born}
	F_{n m}(\mathbf{R}) \rightarrow e^{i \mathbf{k}_{0}\cdot\mathbf{R}}+\frac{e^{i \mathbf{k}\cdot\mathbf{R}}}{\sqrt{\mathbf{R}}}f_{n m}(\theta),
\end{equation}
where it was assumed that the asymptotic form of the solution is the sum of an incoming plane wave and an outgoing circular wave. In Eq. (\ref{eq:born}), $ \mathbf{k}_{0}/\mathbf{k} $ are the initial/final wave vectors for the scattered particles, and 
\begin{equation}\label{eq:angular_f}
		f_{n m}(\theta) =-\frac{Me^{i\pi/4}}{\sqrt{2 \pi k \hbar^{2}}}  \int d\mathbf{r}_{A} d\mathbf{r}_{B}d\mathbf{R}^{\prime}\, e^{i\left(\mathbf{k}_{0}-\mathbf{k}\right) \cdot\mathbf{R}^{\prime}} V\left(\mathbf{r}_{A}, \mathbf{r}_{B}, \mathbf{R}^{\prime}\right)  \chi_{A ,n}^{*}\left(\mathbf{r}_{A}\right)\chi_{A ,1}\left(\mathbf{r}_{A}\right) \chi_{B, m}^{*}\left(\mathbf{r}_{B}\right) \chi_{B ,1}\left(\mathbf{r}_{B}\right).
\end{equation}

The 2D differential scattering cross section is, analogously to that of a 3D system, given by \cite{FENG7096}
\begin{equation}\label{eq:diff_general}
	\begin{aligned}
		I_{n,m}(\theta) &=\frac{k}{k_{0}}\left|f_{n m}(\theta)\right|^{2}\\
		&=\frac{M^{2}}{2 \pi k_{0} \hbar^{4}}  \left|\int d\mathbf{r}_{A} d\mathbf{r}_{B}d\mathbf{R}^{\prime}\, e^{i\mathbf{q}\cdot\mathbf{R}^{\prime}} V\left(\mathbf{r}_{A}, \mathbf{r}_{B}, \mathbf{R}^{\prime}\right)  \chi_{A ,n}^{*}\left(\mathbf{r}_{A}\right)\chi_{A ,1}\left(\mathbf{r}_{A}\right) \chi_{B, m}^{*}\left(\mathbf{r}_{B}\right) \chi_{B ,1}\left(\mathbf{r}_{B}\right)\right|^{2},
	\end{aligned}
\end{equation}
with $ \mathbf{q}=\mathbf{k}_{0}-\mathbf{k} $. 
Finally, recalling that the free carrier (electron or hole) has no internal structure, the sum of internal energies of the system reduces to that of the exciton. Furthermore, Eq. (\ref{eq:diff_general}) can be simplified further by assuming a central field approximation, reading
\begin{equation}\label{eq:diff_general_final}
	I_{n,m}(\theta) =\frac{M^{2}}{2 \pi k_{0} \hbar^{4}}  \left|\int d\mathbf{r} d\mathbf{R}^{\prime}\, e^{i\mathbf{q}\cdot\mathbf{R}^{\prime}} V\left(\mathbf{r}, \mathbf{R}^{\prime}\right)  \chi_{n}^{*}\left(\mathbf{r}\right)\chi_{1}\left(\mathbf{r}\right)\right|^{2},
\end{equation}
where $ \chi $ is now the wave function of the exciton, $ \mathbf{r} $ is the relative position of the electron and hole in the exciton and $ \mathbf{R}^{\prime} $ is the relative position from the free carrier to the center of mass of the exciton.

\section{Computation of Scattering Contribution to Excitonic Linewidth \label{app:numeric_integral}}

We begin by changing the integration limits $\left[0,+\infty\right)$ to a finite limit, in this case $\left[0,1\right]$, via a change of variables defined as $k=\tan\left(\frac{\pi x}{2}\right)$. With this change of variables, the integral of Eq. (\ref{eq:linewidth_tot}) reads 
\begin{equation}\label{eq:linewidth_tot_limits}
	\Gamma_{\mathrm{Total}}=\frac{2\hbar^{2}}{\pi M}\int_{0}^{1}dx\,\frac{dk}{dx}k(x)^{2}n_{F}\left(\frac{m_c+m_e+m_h}{m_e+m_h}k(x)\right)Q_{\mathrm{Total}}(k(x)),
\end{equation}
We can then define a grid of points $ x_i $ for our discretization, meaning that
\begin{equation}\label{eq:linewidth_tot_discrete}
	\Gamma_{\mathrm{Total}}=\frac{2\hbar^{2}}{\pi M}\sum_{i=1}^{N}w_{i}\frac{dk}{dx_{i}}k_{i}^{2}n_{F}\left(\frac{m_c+m_e+m_h}{m_e+m_h}k_{i}\right)Q_{\mathrm{Total}}(k_{i}),
\end{equation}
where $N$ is the number of points considered in the discretization, $w_{i}$ is the weight function of the quadrature in question, and the discretized variables are defined as $q_{i}\equiv q\left(x_{i}\right)$, and $\frac{dk}{dx_{i}}\equiv\left.\frac{dk}{dx}\right|_{x=x_{i}}$. For the numerical quadrature, we employ a Gauss--Legendre quadrature\cite{Kythe2002}, defined as 
\[
\int_{a}^{b}f\left(x\right)dx\approx\sum_{i=1}^{N}f\left(x_{i}\right)w_{i},
\]
where 
\begin{align}
	x_{i}&=\frac{a+b+\left(b-a\right)\xi_{i}}{2}, &w_{i}&=\frac{b-a}{\left(1-\xi_{i}^{2}\right)\left[\left.\frac{dP_{N}\left(x\right)}{dx}\right|_{x=\xi_{i}}\right]^{2}},\nonumber
\end{align}
with $\xi_{i}$ the $i$-th zero of the Legendre polynomial $P_{N}\left(x\right)$.

\begin{adjustwidth}{-\extralength}{0cm}

\reftitle{References}


\bibliographystyle{unsrt}
\bibliography{scatter_biblio}

\begin{thebibliography}{999}

\bibitem[\`a~la Guillaume \em{et~al.}(1969)\`a~la Guillaume, Debever, and
  Salvan]{PhysRev.177.567}
\`a~la Guillaume, C.B.; Debever, J.M.; Salvan, F.
\newblock Radiative Recombination in Highly Excited CdS.
\newblock {\em Phys. Rev.} {\bf 1969}, {\em 177},~567--580.
\newblock {\url{https://doi.org/10.1103/PhysRev.177.567}}.

\bibitem[Iwai and Namba(1971)]{10.1063/1.1653814}
Iwai, S.; Namba, S.
\newblock Emission Spectra in CdS under High Excitation by Electron Beam.
\newblock {\em Applied Physics Letters} {\bf 1971}, {\em 19},~41--43.
\newblock {\url{https://doi.org/10.1063/1.1653814}}.

\bibitem[In~Yu \em{et~al.}(1973)In~Yu, Goto, and Ueta]{10.1143/JPSJ.34.693}
In~Yu, C.; Goto, T.; Ueta, M.
\newblock Emission of Cuprous Halide Crystals at High Density Excitation.
\newblock {\em Journal of the Physical Society of Japan} {\bf 1973}, {\em
  34},~693--698.
\newblock {\url{https://doi.org/10.1143/JPSJ.34.693}}.

\bibitem[Braun \em{et~al.}(1973)Braun, Bille, Fischer, and
  Huber]{10.1002/pssb.2220580237}
Braun, W.; Bille, J.; Fischer, T.; Huber, G.
\newblock Laser Emission in CdSe Due to Exciton—Exciton and
  Exciton—Electron Interaction.
\newblock {\em physica status solidi (b)} {\bf 1973}, {\em 58},~759--765.
\newblock {\url{https://doi.org/https://doi.org/10.1002/pssb.2220580237}}.

\bibitem[Levy and Grun(1974)]{10.1002/pssa.2210220102}
Levy, R.; Grun, J.B.
\newblock Optical properties of strongly excited direct band gap materials.
\newblock {\em physica status solidi (a)} {\bf 1974}, {\em 22},~11--38.
\newblock {\url{https://doi.org/https://doi.org/10.1002/pssa.2210220102}}.

\bibitem[Klingshirn(1975)]{10.1002/pssb.2220710216}
Klingshirn, C.
\newblock The Luminescence of ZnO under High One- and Two-Quantum Excitation.
\newblock {\em physica status solidi (b)} {\bf 1975}, {\em 71},~547--556.
\newblock {\url{https://doi.org/https://doi.org/10.1002/pssb.2220710216}}.

\bibitem[Hönerlage \em{et~al.}(1976)Hönerlage, Klingshirn, and
  Grun]{10.1002/pssb.2220780219}
Hönerlage, B.; Klingshirn, C.; Grun, J.B.
\newblock Spontaneous emission due to exciton—electron scattering in
  semiconductors.
\newblock {\em physica status solidi (b)} {\bf 1976}, {\em 78},~599--608.
\newblock {\url{https://doi.org/https://doi.org/10.1002/pssb.2220780219}}.

\bibitem[Levy \em{et~al.}(1975)Levy, Grun, and
  Nikitine]{Levy1975ExperimentalIO}
Levy, R.; Grun, J.B.; Nikitine, S.
\newblock Experimental investigation of the competition of stimulated emissions
  involving exciton.
\newblock {\em Springer Tracts in Modern Physics} {\bf 1975}, {\em
  73},~211--219.

\bibitem[Elkomoss and Munschy(1977)]{ELKOMOSS1977557}
Elkomoss, S.G.; Munschy, G.
\newblock Electron-exciton elastic scattering cross sections in the central
  field and the exchange approximations.
\newblock {\em Journal of Physics and Chemistry of Solids} {\bf 1977}, {\em
  38},~557--563.
\newblock {\url{https://doi.org/https://doi.org/10.1016/0022-3697(77)90220-7}}.

\bibitem[Elkomoss and Munschy(1979)]{ELKOMOSS1979431}
Elkomoss, S.G.; Munschy, G.
\newblock Electron-exciton inelastic collision cross sections for different
  semiconductors.
\newblock {\em Journal of Physics and Chemistry of Solids} {\bf 1979}, {\em
  40},~431--438.
\newblock {\url{https://doi.org/https://doi.org/10.1016/0022-3697(79)90058-1}}.

\bibitem[Reynolds \em{et~al.}(1986)Reynolds, Bajaj, Litton, Singh, Yu, Pearah,
  Klem, and Morkoc]{Reynolds1986}
Reynolds, D.C.; Bajaj, K.K.; Litton, C.W.; Singh, J.; Yu, P.W.; Pearah, P.;
  Klem, J.; Morkoc, H.
\newblock High-resolution photoluminescence and reflection studies of
  GaAs-${\mathrm{Al}}_{\mathrm{x}}$${\mathrm{Ga}}_{1\mathrm{\ensuremath{-}}\mathrm{x}}$As
  multi-quantum-well structures grown by molecular-beam epitaxy: Determination
  of microscopic structural quality of interfaces.
\newblock {\em Phys. Rev. B} {\bf 1986}, {\em 33},~5931--5934.
\newblock {\url{https://doi.org/10.1103/PhysRevB.33.5931}}.

\bibitem[Museur \em{et~al.}(2008)Museur, Feldbach, and Kanaev]{Museur2008}
Museur, L.; Feldbach, E.; Kanaev, A.
\newblock Defect-related photoluminescence of hexagonal boron nitride.
\newblock {\em Phys. Rev. B} {\bf 2008}, {\em 78},~155204.
\newblock {\url{https://doi.org/10.1103/PhysRevB.78.155204}}.

\bibitem[Cadiz \em{et~al.}(2017)Cadiz, Courtade, Robert, Wang, Shen, Cai,
  Taniguchi, Watanabe, Carrere, Lagarde, Manca, Amand, Renucci, Tongay, Marie,
  and Urbaszek]{Cadiz2017}
Cadiz, F.; Courtade, E.; Robert, C.; Wang, G.; Shen, Y.; Cai, H.; Taniguchi,
  T.; Watanabe, K.; Carrere, H.; Lagarde, D.;  et~al.
\newblock Excitonic Linewidth Approaching the Homogeneous Limit in
  ${\mathrm{MoS}}_{2}$-Based van der Waals Heterostructures.
\newblock {\em Phys. Rev. X} {\bf 2017}, {\em 7},~021026.
\newblock {\url{https://doi.org/10.1103/PhysRevX.7.021026}}.

\bibitem[Elias \em{et~al.}(2019)Elias, Valvin, Pelini, Summerfield, Mellor,
  Cheng, Eaves, Foxon, Beton, Novikov, Gil, and Cassabois]{Elias2019}
Elias, C.; Valvin, P.; Pelini, T.; Summerfield, A.; Mellor, C.J.; Cheng, T.S.;
  Eaves, L.; Foxon, C.T.; Beton, P.H.; Novikov, S.V.;  et~al.
\newblock Direct band-gap crossover in epitaxial monolayer boron nitride.
\newblock {\em Nature Communications} {\bf 2019}, {\em 10},~2639.
\newblock {\url{https://doi.org/10.1038/s41467-019-10610-5}}.

\bibitem[Li \em{et~al.}(2021)Li, Wang, Zhang, Elias, Ye, Evans, Eda, Redwing,
  Cassabois, Gil, Valvin, He, Liu, and Edgar]{Li2021a}
Li, J.; Wang, J.; Zhang, X.; Elias, C.; Ye, G.; Evans, D.; Eda, G.; Redwing,
  J.M.; Cassabois, G.; Gil, B.;  et~al.
\newblock Hexagonal Boron Nitride Crystal Growth from Iron, a Single Component
  Flux.
\newblock {\em ACS Nano} {\bf 2021}, {\em 15},~7032--7039.
\newblock PMID: 33818058, {\url{https://doi.org/10.1021/acsnano.1c00115}}.

\bibitem[Vuong \em{et~al.}(2017)Vuong, Cassabois, Valvin, Rousseau,
  Summerfield, Mellor, Cho, Cheng, Albar, Eaves, Foxon, Beton, Novikov, and
  Gil]{Vuong2017}
Vuong, T.Q.P.; Cassabois, G.; Valvin, P.; Rousseau, E.; Summerfield, A.;
  Mellor, C.J.; Cho, Y.; Cheng, T.S.; Albar, J.D.; Eaves, L.;  et~al.
\newblock Deep ultraviolet emission in hexagonal boron nitride grown by
  high-temperature molecular beam epitaxy.
\newblock {\em 2D Materials} {\bf 2017}, {\em 4},~021023.
\newblock {\url{https://doi.org/10.1088/2053-1583/aa604a}}.

\bibitem[Kim \em{et~al.}(2015)Kim, Hsu, Park, Chae, Yun, Lee, Cho, Fang, Lee,
  Palacios, Dresselhaus, Kim, Lee, and Kong]{Kim2015}
Kim, S.M.; Hsu, A.; Park, M.H.; Chae, S.H.; Yun, S.J.; Lee, J.S.; Cho, D.H.;
  Fang, W.; Lee, C.; Palacios, T.;  et~al.
\newblock Synthesis of large-area multilayer hexagonal boron nitride for high
  material performance.
\newblock {\em Nature Communications} {\bf 2015}, {\em 6},~8662.
\newblock {\url{https://doi.org/10.1038/ncomms9662}}.

\bibitem[Yano \em{et~al.}(2000)Yano, Yap, Okamoto, Onda, Yoshimura, Mori, and
  Sasaki]{Yano2000}
Yano, M.; Yap, Y.K.; Okamoto, M.; Onda, M.; Yoshimura, M.; Mori, Y.; Sasaki, T.
\newblock Na: A New Flux for Growing Hexagonal Boron Nitride Crystals at Low
  Temperature.
\newblock {\em Japanese Journal of Applied Physics} {\bf 2000}, {\em
  39},~L300--L302.
\newblock {\url{https://doi.org/10.1143/jjap.39.l300}}.

\bibitem[Dingle \em{et~al.}(1974)Dingle, Wiegmann, and
  Henry]{PhysRevLett.33.827}
Dingle, R.; Wiegmann, W.; Henry, C.H.
\newblock Quantum States of Confined Carriers in Very Thin
  ${\mathrm{Al}}_{x}{\mathrm{Ga}}_{1\ensuremath{-}x}\mathrm{As}$-GaAs-${\mathrm{Al}}_{x}{\mathrm{Ga}}_{1\ensuremath{-}x}\mathrm{As}$
  Heterostructures.
\newblock {\em Phys. Rev. Lett.} {\bf 1974}, {\em 33},~827--830.
\newblock {\url{https://doi.org/10.1103/PhysRevLett.33.827}}.

\bibitem[Miller \em{et~al.}(1982)Miller, Chemla, Eilenberger, Smith, Gossard,
  and Tsang]{10.1063/1.93648}
Miller, D.A.B.; Chemla, D.S.; Eilenberger, D.J.; Smith, P.W.; Gossard, A.C.;
  Tsang, W.T.
\newblock Large room-temperature optical nonlinearity in
  ${\mathrm{Ga}\mathrm{As}/\mathrm{Ga}_{1\ensuremath{-}x}\mathrm{Al}_{x}\mathrm{As}}$
  multiple quantum well structures.
\newblock {\em Applied Physics Letters} {\bf 1982}, {\em 41},~679--681.
\newblock {\url{https://doi.org/10.1063/1.93648}}.

\bibitem[Poellmann \em{et~al.}(2015)Poellmann, Steinleitner, Leierseder,
  Nagler, Plechinger, Porer, Bratschitsch, Sch{\"u}ller, Korn, and
  Huber]{Poellmann2015}
Poellmann, C.; Steinleitner, P.; Leierseder, U.; Nagler, P.; Plechinger, G.;
  Porer, M.; Bratschitsch, R.; Sch{\"u}ller, C.; Korn, T.; Huber, R.
\newblock Resonant internal quantum transitions and femtosecond radiative decay
  of excitons in monolayer WSe2.
\newblock {\em Nature Materials} {\bf 2015}, {\em 14},~889--893.
\newblock {\url{https://doi.org/10.1038/nmat4356}}.

\bibitem[Merkl \em{et~al.}(2019)Merkl, Mooshammer, Steinleitner, Girnghuber,
  Lin, Nagler, Holler, Sch{\"u}ller, Lupton, Korn, Ovesen, Brem, Malic, and
  Huber]{Merkl2019}
Merkl, P.; Mooshammer, F.; Steinleitner, P.; Girnghuber, A.; Lin, K.Q.; Nagler,
  P.; Holler, J.; Sch{\"u}ller, C.; Lupton, J.M.; Korn, T.;  et~al.
\newblock Ultrafast transition between exciton phases in van der Waals
  heterostructures.
\newblock {\em Nature Materials} {\bf 2019}, {\em 18},~691--696.
\newblock {\url{https://doi.org/10.1038/s41563-019-0337-0}}.

\bibitem[Lee \em{et~al.}(1986)Lee, Koteles, and Vassell]{PhysRevB.33.5512}
Lee, J.; Koteles, E.S.; Vassell, M.O.
\newblock Luminescence linewidths of excitons in GaAs quantum wells below 150
  K.
\newblock {\em Phys. Rev. B} {\bf 1986}, {\em 33},~5512--5516.
\newblock {\url{https://doi.org/10.1103/PhysRevB.33.5512}}.

\bibitem[Spector \em{et~al.}(1986)Spector, Lee, and Melman]{PhysRevB.34.2554}
Spector, H.N.; Lee, J.; Melman, P.
\newblock Exciton linewidth in semiconducting quantum-well structures.
\newblock {\em Phys. Rev. B} {\bf 1986}, {\em 34},~2554--2560.
\newblock {\url{https://doi.org/10.1103/PhysRevB.34.2554}}.

\bibitem[Henriques \em{et~al.}(2021)Henriques, Mortensen, and
  Peres]{PhysRevB.103.235402}
Henriques, J.C.G.; Mortensen, N.A.; Peres, N.M.R.
\newblock Analytical description of the $1s$ exciton linewidth temperature
  dependence in transition metal dichalcogenides.
\newblock {\em Phys. Rev. B} {\bf 2021}, {\em 103},~235402.
\newblock {\url{https://doi.org/10.1103/PhysRevB.103.235402}}.

\bibitem[Cingolani and Ploog(1991)]{10.1080/00018739100101522}
Cingolani, R.; Ploog, K.
\newblock Frequency and density dependent radiative recombination processes in
  III–V semiconductor quantum wells and superlattices.
\newblock {\em Advances in Physics} {\bf 1991}, {\em 40},~535--623.
\newblock {\url{https://doi.org/10.1080/00018739100101522}}.

\bibitem[Palummo \em{et~al.}(2015)Palummo, Bernardi, and
  Grossman]{10.1021/nl503799t}
Palummo, M.; Bernardi, M.; Grossman, J.C.
\newblock Exciton Radiative Lifetimes in Two-Dimensional Transition Metal
  Dichalcogenides.
\newblock {\em Nano Letters} {\bf 2015}, {\em 15},~2794--2800.
\newblock PMID: 25798735, {\url{https://doi.org/10.1021/nl503799t}}.

\bibitem[Singh and Bajaj(1984)]{10.1063/1.94649}
Singh, J.; Bajaj, K.K.
\newblock Theory of excitonic photoluminescence linewidth in semiconductor
  alloys.
\newblock {\em Applied Physics Letters} {\bf 1984}, {\em 44},~1075--1077.
\newblock {\url{https://doi.org/10.1063/1.94649}}.

\bibitem[Basu(1990)]{10.1063/1.102583}
Basu, P.K.
\newblock Linewidth of free excitons in quantum wells: Contribution by alloy
  disorder scattering.
\newblock {\em Applied Physics Letters} {\bf 1990}, {\em 56},~1110--1112.
\newblock {\url{https://doi.org/10.1063/1.102583}}.

\bibitem[Li \em{et~al.}(1995)Li, Wang, Song, Liang, Xu, Zhu, Zheng, Liao, and
  Yang]{10.1063/1.360578}
Li, W.; Wang, Z.; Song, A.; Liang, J.; Xu, B.; Zhu, Z.; Zheng, W.; Liao, Q.;
  Yang, B.
\newblock Photoluminescence studies on very high‐density
  quasi‐two‐dimensional electron gases in pseudomorphic modulation‐doped
  quantum wells.
\newblock {\em Journal of Applied Physics} {\bf 1995}, {\em 78},~593--595.
\newblock {\url{https://doi.org/10.1063/1.360578}}.

\bibitem[Djuri{\v{s}}i{\'{c}} \em{et~al.}(2005)Djuri{\v{s}}i{\'{c}}, Kwok,
  Leung, Chan, Phillips, Lin, and Gwo]{Djurii2005}
Djuri{\v{s}}i{\'{c}}, A.B.; Kwok, W.M.; Leung, Y.H.; Chan, W.K.; Phillips,
  D.L.; Lin, M.S.; Gwo, S.
\newblock Ultrafast spectroscopy of stimulated emission in single {ZnO}
  tetrapod nanowires.
\newblock {\em Nanotechnology} {\bf 2005}, {\em 17},~244--249.
\newblock {\url{https://doi.org/10.1088/0957-4484/17/1/041}}.

\bibitem[Young \em{et~al.}(1990)Young, Wood, and Charbonneau]{Young1990}
Young, J.F.; Wood, B.M.; Charbonneau, S., Optical Probes of Resonant Tunneling
  Structures.
\newblock In {\em Electronic Properties of Multilayers and Low-Dimensional
  Semiconductor Structures}; Springer US: Boston, MA,  1990; pp. 331--349.
\newblock {\url{https://doi.org/10.1007/978-1-4684-7412-1_19}}.

\bibitem[Teran \em{et~al.}(2009)Teran, Mart{\'{\i}}n, Calleja, Vi{\~{n}}a,
  Eaves, and Henini]{Teran2009}
Teran, F.J.; Mart{\'{\i}}n, M.D.; Calleja, J.M.; Vi{\~{n}}a, L.; Eaves, L.;
  Henini, M.
\newblock Carrier injection effects on exciton dynamics in {GaAs}/{AlAs}
  resonant-tunneling diodes.
\newblock {\em {EPL} (Europhysics Letters)} {\bf 2009}, {\em 85},~67010.
\newblock {\url{https://doi.org/10.1209/0295-5075/85/67010}}.

\bibitem[Koh \em{et~al.}(1997)Koh, Feng, and Spector]{10.1063/1.363972}
Koh, T.S.; Feng, Y.P.; Spector, H.N.
\newblock Effect of an electric field on the scattering of excitons by free
  carriers in semiconducting quantum-well structures.
\newblock {\em Journal of Applied Physics} {\bf 1997}, {\em 81},~2704--2708.
\newblock {\url{https://doi.org/10.1063/1.363972}}.

\bibitem[Vella \em{et~al.}(2022)Vella, Barbosa, Trevisanutto, Verzhbitskiy,
  Zhou, Watanabe, Taniguchi, Kajikawa, and Eda]{10.1002/adom.202102132}
Vella, D.; Barbosa, M.B.; Trevisanutto, P.E.; Verzhbitskiy, I.; Zhou, J.Y.;
  Watanabe, K.; Taniguchi, T.; Kajikawa, K.; Eda, G.
\newblock In-Plane Field-Driven Excitonic Electro-Optic Modulation in Monolayer
  Semiconductor.
\newblock {\em Advanced Optical Materials} {\bf 2022}, {\em 10},~2102132.
\newblock {\url{https://doi.org/https://doi.org/10.1002/adom.202102132}}.

\bibitem[Efimkin \em{et~al.}(2021)Efimkin, Laird, Levinsen, Parish, and
  MacDonald]{PhysRevB.103.075417}
Efimkin, D.K.; Laird, E.K.; Levinsen, J.; Parish, M.M.; MacDonald, A.H.
\newblock Electron-exciton interactions in the exciton-polaron problem.
\newblock {\em Phys. Rev. B} {\bf 2021}, {\em 103},~075417.
\newblock {\url{https://doi.org/10.1103/PhysRevB.103.075417}}.

\bibitem[Feng and Spector(1987)]{FENG1987593}
Feng, Y.P.; Spector, H.N.
\newblock Scattering of excitons by free carriers in semiconducting quantum
  well structures.
\newblock {\em Journal of Physics and Chemistry of Solids} {\bf 1987}, {\em
  48},~593--601.
\newblock {\url{https://doi.org/https://doi.org/10.1016/0022-3697(87)90146-6}}.

\bibitem[Massey and Moiseiwitsch(1951)]{Massey1951}
Massey, H.S.W.; Moiseiwitsch, B.L.
\newblock The application of variational methods to atomic scattering problems
  - I. The elastic scattering of electrons by hydrogen atoms.
\newblock {\em Proceedings of the Royal Society of London. Series A.
  Mathematical and Physical Sciences} {\bf 1951}, {\em 205},~483--496.
\newblock {\url{https://doi.org/10.1098/rspa.1951.0044}}.

\bibitem[Bates and Griffing(1953)]{Bates1953}
Bates, D.R.; Griffing, G.
\newblock Inelastic Collisions between Heavy Particles I: Excitation and
  Ionization of Hydrogen Atoms in Fast Encounters with Protons and with other
  Hydrogen Atoms.
\newblock {\em Proceedings of the Physical Society. Section A} {\bf 1953}, {\em
  66},~961--971.
\newblock {\url{https://doi.org/10.1088/0370-1298/66/11/301}}.

\bibitem[Mott and Massey(1965)]{Mott1965}
Mott, N.F.; Massey, H.S.W.
\newblock {\em Theory of atomic collisions}, 3 ed.; Monographs on Physics,
  Oxford University Press: London, England,  1965.

\bibitem[Feng and Spector(1988)]{FENG7096}
Feng, Y.P.; Spector, H.
\newblock Scattering of screened excitons by free carriers in semiconducting
  quantum well structures.
\newblock {\em IEEE Journal of Quantum Electronics} {\bf 1988}, {\em
  24},~1659--1663.
\newblock {\url{https://doi.org/10.1109/3.7096}}.

\bibitem[Bak and Newman(2010)]{Bak2010}
Bak, J.; Newman, D.J.
\newblock {\em Complex Analysis}; Springer New York,  2010.
\newblock {\url{https://doi.org/10.1007/978-1-4419-7288-0}}.

\bibitem[Rytova((1967))]{rytova1967}
Rytova, S.N.
\newblock The Screened Potential of a Point Charge in a Thin Film.
\newblock {\em Mosc. Un. Phys. Bul.} {\bf (1967)}, {\em 22}.

\bibitem[Keldysh((1979))]{keldysh1979coulomb}
Keldysh, L.V.
\newblock Coulomb interaction in thin semiconductor and semimetal films.
\newblock {\em Sov. J. Exp. and Theor. Phys. Lett.} {\bf (1979)}, {\em
  29},~658.

\bibitem[Li and Appelbaum((2019))]{PhysRevB.99.035429}
Li, P.; Appelbaum, I.
\newblock Excitons without effective mass: Biased bilayer graphene.
\newblock {\em Phys. Rev. B} {\bf (2019)}, {\em 99},~035429.
\newblock {\url{https://doi.org/10.1103/PhysRevB.99.035429}}.

\bibitem[Quintela \em{et~al.}(2022)Quintela, Henriques, Tenório, and
  Peres]{Quintela}
Quintela, M.F.C.M.; Henriques, J.C.G.; Tenório, L.G.M.; Peres, N.M.R.
\newblock Theoretical Methods for Excitonic Physics in 2D Materials.
\newblock {\em physica status solidi (b)} {\bf 2022}, {\em n/a},~2200097.
\newblock {\url{https://doi.org/https://doi.org/10.1002/pssb.202200097}}.

\bibitem[Gomes \em{et~al.}(2021)Gomes, Trallero-Giner, and
  Vasilevskiy]{Gomes2021}
Gomes, J.N.S.; Trallero-Giner, C.; Vasilevskiy, M.I.
\newblock Variational calculation of the lowest exciton states in phosphorene
  and transition metal dichalcogenides.
\newblock {\em Journal of Physics: Condensed Matter} {\bf 2021}, {\em
  34},~045702.

\bibitem[Lee and Lin(1979)]{Lee1979}
Lee, Y.C.; Lin, D.L.
\newblock Wannier excitons in a thin crystal film.
\newblock {\em Phys. Rev. B} {\bf 1979}, {\em 19},~1982--1989.
\newblock {\url{https://doi.org/10.1103/PhysRevB.19.1982}}.

\bibitem[Yang \em{et~al.}(1991)Yang, Guo, Chan, Wong, and
  Ching]{PhysRevA.43.1186}
Yang, X.L.; Guo, S.H.; Chan, F.T.; Wong, K.W.; Ching, W.Y.
\newblock Analytic solution of a two-dimensional hydrogen atom. I.
  Nonrelativistic theory.
\newblock {\em Phys. Rev. A} {\bf 1991}, {\em 43},~1186--1196.
\newblock {\url{https://doi.org/10.1103/PhysRevA.43.1186}}.

\bibitem[Wannier(1937)]{Wannier1937}
Wannier, G.H.
\newblock The Structure of Electronic Excitation Levels in Insulating Crystals.
\newblock {\em Phys. Rev.} {\bf 1937}, {\em 52},~191--197.

\bibitem[Serway \em{et~al.}(2003)Serway, Faughn, and Moses]{Serway2003}
Serway, R.A.; Faughn, J.S.; Moses, C.J.
\newblock {\em College Physics}; Number vol. 1 in College Physics, Brooks/Cole,
   2003.

\bibitem[Ferreira \em{et~al.}(2019)Ferreira, Chaves, Peres, and
  Ribeiro]{Ferreira19}
Ferreira, F.; Chaves, A.J.; Peres, N.M.R.; Ribeiro, R.M.
\newblock Excitons in hexagonal boron nitride single-layer: a new platform for
  polaritonics in the ultraviolet.
\newblock {\em J. Opt. Soc. Am. B} {\bf 2019}, {\em 36},~674--683.
\newblock {\url{https://doi.org/10.1364/JOSAB.36.000674}}.

\bibitem[Henriques \em{et~al.}(2020)Henriques, Ventura, Fernandes, and
  Peres]{henriques_optical_2020}
Henriques, J.C.G.; Ventura, G.B.; Fernandes, C.D.M.; Peres, N.M.R.
\newblock Optical absorption of single-layer hexagonal boron nitride in the
  ultraviolet.
\newblock {\em J. of Phys.: Cond. Matt.} {\bf 2020}, {\em 32},~025304.

\bibitem[Selig \em{et~al.}(2016)Selig, Bergh{\"a}user, Raja, Nagler,
  Sch{\"u}ller, Heinz, Korn, Chernikov, Malic, and Knorr]{Selig2016}
Selig, M.; Bergh{\"a}user, G.; Raja, A.; Nagler, P.; Sch{\"u}ller, C.; Heinz,
  T.F.; Korn, T.; Chernikov, A.; Malic, E.; Knorr, A.
\newblock Excitonic linewidth and coherence lifetime in monolayer transition
  metal dichalcogenides.
\newblock {\em Nature Communications} {\bf 2016}, {\em 7},~13279.
\newblock {\url{https://doi.org/10.1038/ncomms13279}}.

\bibitem[Palummo \em{et~al.}(2015)Palummo, Bernardi, and Grossman]{Palummo2015}
Palummo, M.; Bernardi, M.; Grossman, J.C.
\newblock Exciton Radiative Lifetimes in Two-Dimensional Transition Metal
  Dichalcogenides.
\newblock {\em Nano Letters} {\bf 2015}, {\em 15},~2794--2800.
\newblock PMID: 25798735, {\url{https://doi.org/10.1021/nl503799t}}.

\bibitem[Henriques \em{et~al.}(2021)Henriques, Mortensen, and
  Peres]{Henriques2021}
Henriques, J.C.G.; Mortensen, N.A.; Peres, N.M.R.
\newblock Analytical description of the $1s$ exciton linewidth temperature
  dependence in transition metal dichalcogenides.
\newblock {\em Phys. Rev. B} {\bf 2021}, {\em 103},~235402.
\newblock {\url{https://doi.org/10.1103/PhysRevB.103.235402}}.

\bibitem[Feng and Spector(1987)]{FENG1987459}
Feng, Y.P.; Spector, H.N.
\newblock Exciton linewidth due to scattering from free carriers in
  semiconducting quantum well structures.
\newblock {\em Superlattices and Microstructures} {\bf 1987}, {\em
  3},~459--461.
\newblock {\url{https://doi.org/https://doi.org/10.1016/0749-6036(87)90223-0}}.

\bibitem[Feng(1989)]{feng1989excitons}
Feng, Y.P.
\newblock Excitons in semiconducting quantum well structures.
\newblock PhD thesis,  1989.

\bibitem[Koh \em{et~al.}(1997)Koh, Feng, and Spector]{10.1063/1.364274}
Koh, T.S.; Feng, Y.P.; Spector, H.N.
\newblock Exciton linewidth due to scattering by free carriers in
  semiconducting quantum well structures: Finite confining potential model.
\newblock {\em Journal of Applied Physics} {\bf 1997}, {\em 81},~2236--2240.
\newblock {\url{https://doi.org/10.1063/1.364274}}.

\bibitem[Kythe and Puri((2002))]{Kythe2002}
Kythe, P.K.; Puri, P.
\newblock {\em Computational Methods for Linear Integral Equations};
  Birkh\"{a}user Boston,  (2002).
\newblock {\url{https://doi.org/10.1007/978-1-4612-0101-4}}.

\bibitem[Lampert(1958)]{Lampert1958}
Lampert, M.A.
\newblock Mobile and Immobile Effective-Mass-Particle Complexes in Nonmetallic
  Solids.
\newblock {\em Phys. Rev. Lett.} {\bf 1958}, {\em 1},~450--453.
\newblock {\url{https://doi.org/10.1103/PhysRevLett.1.450}}.

\bibitem[Kheng \em{et~al.}(1993)Kheng, Cox, d'~Aubign\'e, Bassani, Saminadayar,
  and Tatarenko]{Kheng1993}
Kheng, K.; Cox, R.T.; d'~Aubign\'e, M.Y.; Bassani, F.; Saminadayar, K.;
  Tatarenko, S.
\newblock Observation of negatively charged excitons
  ${\mathit{X}}^{\mathrm{\ensuremath{-}}}$ in semiconductor quantum wells.
\newblock {\em Phys. Rev. Lett.} {\bf 1993}, {\em 71},~1752--1755.
\newblock {\url{https://doi.org/10.1103/PhysRevLett.71.1752}}.

\bibitem[Chen \em{et~al.}(2002)Chen, Stievater, Batteh, Li, Steel, Gammon,
  Katzer, Park, and Sham]{Chen2002}
Chen, G.; Stievater, T.H.; Batteh, E.T.; Li, X.; Steel, D.G.; Gammon, D.;
  Katzer, D.S.; Park, D.; Sham, L.J.
\newblock Biexciton Quantum Coherence in a Single Quantum Dot.
\newblock {\em Phys. Rev. Lett.} {\bf 2002}, {\em 88},~117901.
\newblock {\url{https://doi.org/10.1103/PhysRevLett.88.117901}}.

\bibitem[Li \em{et~al.}(2003)Li, Wu, Steel, Gammon, Stievater, Katzer, Park,
  Piermarocchi, and Sham]{Li2003}
Li, X.; Wu, Y.; Steel, D.; Gammon, D.; Stievater, T.H.; Katzer, D.S.; Park, D.;
  Piermarocchi, C.; Sham, L.J.
\newblock An All-Optical Quantum Gate in a Semiconductor Quantum Dot.
\newblock {\em Science} {\bf 2003}, {\em 301},~809--811.
\newblock {\url{https://doi.org/10.1126/science.1083800}}.

\bibitem[Du \em{et~al.}(2016)Du, Neal, Zhou, and Ye]{Du2016}
Du, Y.; Neal, A.T.; Zhou, H.; Ye, P.D.
\newblock Transport studies in 2D transition metal dichalcogenides and black
  phosphorus.
\newblock {\em Journal of Physics: Condensed Matter} {\bf 2016}, {\em
  28},~263002.
\newblock {\url{https://doi.org/10.1088/0953-8984/28/26/263002}}.

\bibitem[Kuiri \em{et~al.}(2021)Kuiri, Srivastav, Ray, Watanabe, Taniguchi,
  Das, and Das]{Kuiri2021}
Kuiri, M.; Srivastav, S.K.; Ray, S.; Watanabe, K.; Taniguchi, T.; Das, T.; Das,
  A.
\newblock Enhanced electron-phonon coupling in doubly aligned hexagonal boron
  nitride bilayer graphene heterostructure.
\newblock {\em Phys. Rev. B} {\bf 2021}, {\em 103},~115419.
\newblock {\url{https://doi.org/10.1103/PhysRevB.103.115419}}.

\bibitem[Li \em{et~al.}(2021)Li, Goryca, Yumigeta, Li, Tongay, and
  Crooker]{Li2021}
Li, J.; Goryca, M.; Yumigeta, K.; Li, H.; Tongay, S.; Crooker, S.A.
\newblock Valley relaxation of resident electrons and holes in a monolayer
  semiconductor: Dependence on carrier density and the role of
  substrate-induced disorder.
\newblock {\em Phys. Rev. Materials} {\bf 2021}, {\em 5},~044001.
\newblock {\url{https://doi.org/10.1103/PhysRevMaterials.5.044001}}.

\end{thebibliography}

\end{adjustwidth}
\end{document}